\documentclass[12pt,epsf]{article}
\newlength{\abstractwidth}
\usepackage[dvips]{graphicx}
\usepackage{subfigure}

\renewcommand{\thefootnote}{\fnsymbol{footnote}}
\renewcommand{\thanks}[1]{\footnote{#1}} 
\newcommand{\starttext}{
\setcounter{footnote}{0}
\renewcommand{\thefootnote}{\arabic{footnote}}}
\renewcommand{\theequation}{\thesection.\arabic{equation}}
\newcommand{\be}{\begin{equation}}
\newcommand{\bea}{\begin{eqnarray}}
\newcommand{\eea}{\end{eqnarray}}
\newcommand{\beq}{\begin{equation}}
\newcommand{\ee}{\end{equation}}
\newcommand{\eeq}{\end{equation}}

\def\<{\langle}

\renewcommand{\>}{\rangle}
\def\ba{\begin{eqnarray}}
\def\ea{\end{eqnarray}}

\def\14{{1\over4}}
\def\12{{1 \over 2}}
\def\eq{&=&}

\def\h3{h^{3\over 2}}
\def\R{\bar{R}}

\def\>{\rangle}
\def\<{\langle}
\def\sc {Schwarzschild}

\def\0cc{$\Lambda = 0$}

\def\R{R_{ADS}}
\def\t{t_*}
\def\dof{degrees of freedom}
\def\ln{\log{N}}
\def\r{\rho}

\begin{document}
\renewcommand{\theequation}{\thesection.\arabic{equation}}
\begin{titlepage}
\bigskip
\rightline{SU-ITP-08/18}
\rightline{OIQP-08-08}

\bigskip\bigskip\bigskip\bigskip

\centerline{\Large \bf {Fast Scramblers}}

\bigskip\bigskip
\bigskip\bigskip

\centerline{{\it  Yasuhiro Sekino}${}^{1,2}$,  {\it L. Susskind}${}^{2}$  }
\medskip
\centerline{${}^1\;$Okayama Institute for Quantum Physics, Okayama 700-0015, Japan}
\centerline{${}^2\;$Physics Department, Stanford University, Stanford, CA
 94305-4060, USA}
\medskip
\medskip

\bigskip\bigskip
\begin{abstract}
We consider the problem of how fast a quantum system can scramble (thermalize) information, given that the interactions are between bounded clusters of degrees of freedom;  pairwise interactions would be an example. Based on previous work, we conjecture:

\bigskip

 1) The most rapid scramblers take a time logarithmic in the number of degrees of freedom.

 \bigskip

 2)Matrix quantum mechanics (systems whose degrees of freedom are $n$ by
 $n$ matrices) saturate the bound.

 \bigskip

 3) Black holes are the fastest scramblers in nature.

 \bigskip

 The conjectures are based on two sources, one from quantum information theory, and the other from the study of black holes in String Theory.

\end{abstract}

\end{titlepage}
\starttext \baselineskip=18pt \setcounter{footnote}{0}

 \setcounter{equation}{0}
\section{Complementarity}
The principle of Black Hole Complementarity \cite{uglum} could be falsified by exhibiting that some
observer detects a violation of a law of physics.
In order to illustrate what is at stake let us review the ``Alice and
Bob" thought-experiment described in \cite{Larus} \footnote{This
experiment was originally proposed to one of us (L.S.) by John Preskill in 1993.}.

Bob hovers outside a massive Schwarzschild black hole while Alice jumps through the horizon carrying a quantum-bit of information.   Bob then collects the subsequent Hawking photons, and in principle he can reconstruct Alice's bit. Once Bob has collected that information he jumps into the black hole. Meanwhile, Alice sends her q-bit to Bob in the form of a photon. Thus Bob gets to observe quantum cloning of Alice's original q-bit. This of course would contradict the principles of quantum mechanics.

The way out of the paradox involves knowing how long it takes for Bob to collect the bit from the Hawking radiation. If it takes too long then Alice's message will not reach Bob before Bob runs into the singularity. Thus it is important to know how long it takes for the black hole to emit Alice's bit.

The answer, originally suggested in the work of Page \cite{page} is that it takes at least the time necessary to radiate half the entropy of the black hole. In Planck units this is a time of order $M^3$ and is easily long enough to insure that Bob can never observe cloning \cite{Larus}.

The experiment is easily analyzed in the near-horizon limit of Kruskal coordinates.
The near-horizon geometry is flat and can be described by coordinates $X^+, X^-, x^i$ where $X^{\pm}$  are light-like and $x^i$ lie in the tangent plane of the horizon.  The black hole singularity is given by $X^+X^- = R^2$ where $R$ is the \sc \ radius.

\begin{figure}
\begin{center}
\includegraphics{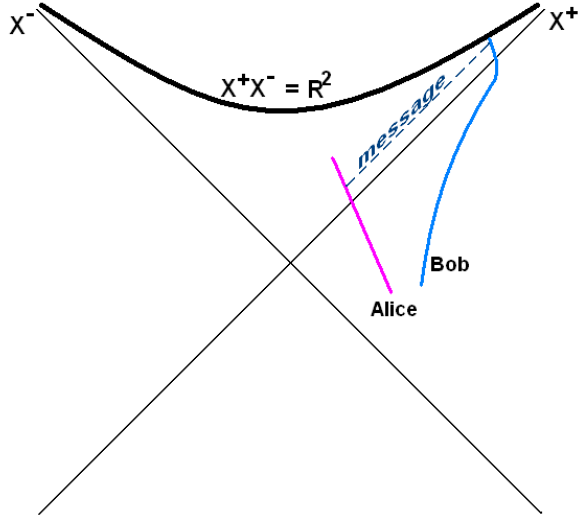}
\caption{A thought experiment in which Bob observes quantum cloning.}
\label{foodah}
\end{center}
\end{figure}

The exterior of the black hole can be described by Rindler coordinates $\rho, \ \omega$ where $\rho$ is proper radial distance from the horizon and $\omega$ is the dimensionless Rindler time. The connection between the Rindler coordinates and $X^{\pm}$ is,
\be
X^{\pm} = \pm \rho e^{\pm \omega}.
\label{01}
\ee
Finally we note that the conventional Schwarzschild time is given by
\be
t=  2 R \omega 
\label{02}
\ee

Now consider Bob who is located outside the black hole, let's say at a distance of order $R$ from the horizon, collecting photons. After a time $t_{retrieval}$ Bob retrieves Alice's bit and jumps in. Equivalently we can define a Rindler retrieval time, $\omega_{ret} = t_{retrieval}/2R$. Bob's jumping-off point occurs at Kruskal coordinate $X^+_{jump}$ given by
\be
X^+_{jump} = R \exp{(\omega_{ret})}.
\label{03}
\ee

From the fact that the singularity is at $X^+ X^- = R^2$ (and that Bob's trajectory is time-like) we deduce that Bob must hit the singularity at a value of $X^-$ satisfying 
\be
X^- < R \exp{(-\omega_{ret})}.
\label{04}
\ee

Now let's go back to Alice. If Bob is to receive the message, Alice must send it during the brief time interval between the time she crosses horizon at $X^- =0$, and the time she crosses  $X^- =R \exp{(-\omega_{ret})}$. In classical physics this would not constrain the energy carried by the message, but  the Uncertainty Principle requires the message to be carried by one or more photons with energy
\be
E_{photon} \sim R^{-1} \exp{(\omega_{ret})}.
\label{05}
\ee

 Alice obviously cannot send the message unless she has that much energy available. But on the other hand she cannot fit in the black hole unless her available energy is much less than the mass of the black hole.  Thus (in Planck units)  $E_{photon} $ must be less than $R$. The upshot is that if Bob is to receive Alice's message, the retrieval time must satisfy
\be
 \exp{(\omega_{ret})} < R^2
 \label{06}
\ee
or 
\be
{\omega_{ret}} < \log R.
\label{07}
\ee

Now the point is that if Bob can receive Alice's message, he has observed cloning and that would violate the principles of quantum mechanics. To put it another way, to guarantee that Bob's observations do not violate quantum principles, we should have
\be
{\omega_{ret}} > \log R.
\label{08}
\ee

Based on Page's work, Susskind and Thorlacius \cite{Larus} assumed that Bob would need to collect at least half the photons that the black hole emits before he  captures Alice's bit, and that takes a Rindler time of order $R^2$ which greatly exceeds the bound (\ref{08}). Evidently the information-cloning principle is protected by a huge ``overkill" factor.

Now we come to the observation of  \cite{patrick}. Hayden and Preskill consider a new twist in which the black hole has been evaporating for a long time before Alice and her bit fall in: so long that more than half the initial entropy has been radiated. Assuming Bob can efficiently use the information in those photons,  then he only needs a small amount of time to capture Alice's bit after she falls in. This version of the experiment presents a much greater danger of observable cloning.

How much time does it take for Alice's bit to be emitted under the new conditions in which half the entropy of the black hole has been emitted before Alice jumps in? The answer derived in \cite{patrick} is a new time-scale that we will call the ``scrambling time." 

To define the scrambling time consider a complex chaotic system of many degrees of freedom, that has originally been prepared in some pure state. After a long time the system thermalizes although its quantum state remains pure. What do we mean that it thermalizes if the state remains pure? Suppose the system has $N$ degrees of freedom. Now consider the density matrix of a subsystem of $m<< N$ degrees of freedom. It is well known that the small subsystem's density matrix will tend toward thermal equilibrium with an average energy given by appropriately partitioning the original average energy of the big system. In other words the entanglement entropy of the subsystem will approach the maximal value.

In fact the subsystem does not have to be very small. The work of Page \cite{page} makes it very plausible that the subsystem will be extremely close to thermal for any $m$ less than
$ N/2$. When this condition is achieved, i.e., when any subsystem smaller than half the whole system has maximum entanglement entropy,  we will call the system ``scrambled." Intuitively it means that any information contained in the original state is mixed up so thoroughly that it can only be recovered by studying at least half the degrees of freedom.

Let us start with a scrambled system and add  a single degree of freedom in a pure state. Alternately we could perturb a small collection of  degrees of freedom. The  system will  no longer be completely scrambled since one can recover information  by looking at a single degree of freedom. But if we wait, the bit of added information will eventually diffuse over all the degrees of freedom and the  system will return to a scrambled state. The time needed to re-scramble when a bit is added is defined to be the scrambling time. Call it $t_*$ (or in Rindler units $\omega_*$).
(The scrambling time defined in this way is not completely precise since we need to specify a precision in how close the subsystems' entropies are to maximal. In this paper we will ignore this complication.)

The remarkable result of Hayden and Preskill \cite{patrick} is simply this: under the right circumstances, the retrieval time for an evaporating system can be as small as the scrambling time. In particular if Bob collects many photons before Alice adds her bit to the system, the retrieval time and the scrambling time are about the same:
\be  
\omega_{ret} \approx \omega_*.
\label{09}
\ee
As Hayden and Preskill explain, Black Hole Complementarity  would be in trouble if $\omega_*$ is less than order $\log R$.

 \setcounter{equation}{0}
\section{Scrambling}
 We assume a conventional quantum system governed by a Hamiltonian $H$. The degrees of freedom  may be  commuting, i.e., bosonic, or anticommuting (spins can be mapped to anticommuting variables), or some combination of both. The total number of \dof\ scales with a parameter $N$. The Hamiltonian consists of terms, each of which involves clusters of at most $m_0$ \dof\ where $m_0$ is a fixed number throughout. The question is: for such a system what is the smallest that the scrambling time can be.

Let us consider a number of examples. Suppose the \dof\ are arranged in a $d$ dimensional periodic array so that each degree of freedom interacts with only a few near neighbors. The linear dimension of the system is proportional to $N^{1 \over d}$. In this case  the time for a signal to propagate from a single cluster to the most distant cluster obviously grows  with $N$ at a rate that satisfies
\be
t_* \ge c \ \ N^{1 \over d}
\label{11}
\ee
where $c$ is a coefficient that does not depend on $N$.

In many examples the effective rate of interaction is temperature dependent. Thus the coefficient $c$ depends on $\beta$. A convenient parameterization is
\be
\tau \equiv {\t \over \beta} \ge  C(\beta)  N^{1 \over d}
\label{12}
\ee
where $C$ is dimensionless.

In most examples that we are aware of thermalization is a process of
diffusion in which the initial perturbation spreads in space to a
distance of order $\sqrt{t}\, $. In that case the bound becomes
\be
\tau \equiv {\t \over \beta} \ge  C(\beta)  N^{2 \over d}.
\label{13}
\ee

Now let us eliminate the restrictions implied by finite dimensionality. In other words we allow arbitrary interactions between any \dof\ as long as the individual interaction terms involve no more than $m_0$ of them. Roughly speaking, we are going to the limit of infinite dimension. Our first conjecture is that (\ref{13}) is replaced by
\be
\tau \ge  C(\beta) \log N.
\label{14}
\ee
We call systems that saturate (\ref{13}) and (\ref{14}) ``fast scramblers.''

 \setcounter{equation}{0}
\section{Matrix Quantum Mechanics}
The  \dof\ of a $U(n)$ symmetric matrix system are the $n \times n$
components of a finite set of traceless hermitian matrices $M^a_{ij}$,
and  their total number  $N$,  is proportional to $n^2$. The example that we will study is called M(atrix) theory because of its close connection with M-theory (a non-perturbative form of String Theory)~\cite{matrix}. Most likely there is nothing unique about it other than we have some String Theory tools to analyze it. In M(atrix) theory the index $a$ runs from $1$ to $9$. There are additional fermionic matrix variables which probably are not important to the finite temperature behavior.

The bosonic part of the Lagrangian has the form,
\be
L = {\rm Tr}  \sum_a \dot{M^a}\dot{M^a} -  {\rm Tr} \sum_{ab}  [M^a, M^b]^2.
\label{21}
\ee
Note that the kinetic terms involve only pairs of \dof\ while each quartic term involves at most four. Thus the condition that the individual terms in $H$ involve finite clusters is satisfied, in this case with clusters of maximal size $4$. Moreover every matrix element $M^a_{ij}$ is coupled to every other $M^b_{kl}$ by at least one quartic term in the Hamiltonian.

We will assume that the perturbations that disturb the system away from
equilibrium involve a small number of \dof\ and typically have the form
of a trace of a product of 
matrices. 
It follows that  they live in a small subgroup of the $U(n)$ degrees of freedom. Later in the paper we will use String dualities to show that M(atrix) theory is a fast scrambler.

Recently, matrix models for black holes were studied with a 
motivation of understanding the issue of information loss. 
In interesting papers \cite{matrixmodels}, the authors observes
the exponential decrease of the large-$n$ thermal correlators 
in the long time limit, which indicates information loss. They
discuss how the finite-$n$ effect modifies this result.
Our perspective is somewhat different from theirs, in the sense
that we focus on properties of pure states 
rather than describing the black hole as a thermal system. 
At the moment we do not know a precise connection between
their analysis and ours, but we think the simple matrix models
of~\cite{matrixmodels} would be useful when we 
try to test our conjecture in the future.

Before discussing M(atrix) theory further, we will describe a non-Hamiltonian example based on random unitary operations acting on pairs of q-bits.

\section{Evidence from Quantum Circuits}
The simplest quantum circuit involving $N$ q-bits is constructed as follows. Time is divided into intervals and in each interval a pair of q-bits are selected at random and allowed to 
``scatter'' by means of a randomly \footnote{Random in this context means random but definite. The randomness is not averaged over. It is the temporal analog of quenched randomness.} chosen
$U(4)$ operator. The number of time-steps is called the depth of the circuit. The circuit acts on any input state of the $N$ q-bits and unitarily transforms it to an output state.

It is known that this system scrambles  in a number of steps that
increases with $N$ like $N \log{N}$ (See \cite{patrick} for references)
\footnote{This result has been shown so far for a special model 
where an element of a finite subgroup of U(4) is 
applied at each step~\cite{dankert}. See~\cite{harrow} for the study of
more general cases and for subtleties of the definition of scrambling.}.

Hayden and Preskill have suggested the following closely related
parallel processing model for a maximally efficient thermalizer. We call
it parallel processing because multiple disjoint pairs are allowed to
interact simultaneously. We will call the time between steps $\beta$,
(keeping in mind that this will roughly correspond 
to the inverse temperature in Hamiltonian systems). 
In the example every q-bit interacts once in each time step.

 Each step begins by randomly pairing the q-bits into  $N/2$ pairs. Any q-bit may pair with any other q-bit but none interact with more than one other. Next we pick $N/2$ random $U(4)$ matrices and allow the q-bit pairs to scatter.  As before the total number of $U(4)$ operations required to scramble the system is $N \log{N}$, but now the parallel processing assembles them into only $\log{N}$ time steps taking total time $t_* = \beta \log{N}$. In the notation of (\ref{14})
 \be
 \tau = {t_*  \over \beta } = C \ln
 \label{31}
 \ee
 with $C$ being independent of $\beta$ in this case.

The precise definition of scrambled is technical. A simple definition in
the q-bit model is that the final state has been randomized with respect
to the Haar measure over the entire $2^N$ dimensional Hilbert space. 
But such randomization is known to be inefficient; it requires a
non-polynomial number of time steps. 
Hayden and Preskill rely on a weaker definition of scrambling that requires only quadratic functions of the density matrix elements to completely randomize, i.e., approach their Haar-scrambled values. With that definition, scrambling takes place on a time scale of order $\log N$ and not smaller. The result does not depend on the assumption of two-body interactions. As long as the number of q-bits in the elementary operations is finite, the minimum scrambling time grows like $\log N$.

The logarithmic growth of $t_*$ is not too surprising. Suppose we fix the state of the first q-bit in some manner. Then after one time step that q-bit has influenced two q-bits, namely itself and the one that it interacted with. After $n$ time steps the first q-bit has influenced $2^n$ q-bits. Obviously the system is not completely scrambled until that first q-bit has influenced all the others. Thus the scrambling time cannot be smaller than order $\log{N}$. That it is that small shows how efficient a scrambler the circuit is.

The system that Hayden and Preskill study is not a conventional Hamiltonian system. It is defined by repeated discrete random unitary operations taking place in parallel. But it is probably not very different from conventional hamiltonian systems such as  M(atrix)
theory. As we will see this expectation is borne out.

An interesting question concerning the relation between the discrete models and the continuous hamiltonian evolution, is what time scale in the latter case corresponds to a single step in the discrete theory. The answer obviously depends on the state of the system. Increasing the energy or temperature speeds things up. We believe the right answer is that the discrete time steps should be identified with time intervals of order
\be
\delta t \sim \epsilon^{-1}
\label{32}
\ee
where $\epsilon$ is the energy per degree of freedom. In many cases it is proportional to the temperature. The time scale (\ref{32})  is the time interval during which every degree of freedom interacts about once: hence its identification with the discrete time steps in the parallel processing circuit.

There is another definition of scrambling that is suggested by the work of Don Page~\cite{page}. Consider \it any \rm subsystem of $m$ q-bits with $m< N/2$, in other words a subsystem less than half the size of the whole system. Page has shown that the entanglement entropy on the subsystem is close to maximal in a Haar-scrambled state. In fact the entropy differs from maximal by less than a single  bit even if the subsystem is just a little smaller than $N/2$. Quantitatively
\be
S_m= m - O(e^{2m-N})
\label{33}
\ee
Any state that satisfies (\ref{33}) for all $m < N/2$
we will call Page-scrambled.

Page has shown the Haar-scrambled implies Page-scrambled, but converse is not true, namely, Page-scrambled does not imply Haar-scrambled. In particular, the scrambler described by Hayden and Preskill is sufficient to Page-scramble despite the fact that it only takes $N \ln$ operations.

 \setcounter{equation}{0}
\section{Evidence from Black Holes}
The new ingredient in this paper is to use the duality between 11 dimensional supergravity and M(atrix) Theory to estimate the scrambling time using the theory of D0-brane black holes. By compactifying the 11D theory on a small circle the theory reduces to a 10D theory of D0-branes. One can consider a sector of the theory with a total D0-brane charge equal to $n$ (As we will see, the number of degrees of freedom, $N$, is not $n$ but rather $n^2$ ). The ground state of this sector is a threshold bound state of the D0-branes, which from the 11D standpoint represents a graviton with $n$ units of Kaluza Klein momentum.

Taking the 10D viewpoint, the ground state can be excited to temperature $T$ and becomes a black hole with a conventional Rindler-type horizon. The black hole is classically stable but quantum-mechanically it slowly decays by Hawking radiation.

The main point for our purposes is that the $n$ D0-brane system is
described by M(atrix)theory, and the black hole is merely the thermally
excited system. Thus, in order to confirm the conjecture that Matrix
systems thermalize in a time of order $\log n$ we are led to ask how
long it takes for information to spread over the horizon of a 10D black
hole with D0-brane charge $n$. For this we can use the ``stretched
horizon'' paradigm~\cite{kip, james}.

\subsection{Information Spreading}
Black holes have no hair, at least when they get old. 
We are interested in how long it takes the black hole to become bald. In other words how long does it take for a bit information to diffuse over the entire horizon?
 The simplest situation is a localized perturbation created on the stretched horizon, thereby disturbing the thermal equilibrium. The perturbation then spreads out until it uniformly covers the stretched horizon. Although we have no mathematical proof, it seems reasonable to identify that time with the scrambling time.
 
 A number of possible perturbations come to mind. One could drop a mass onto the black hole and watch the energy and temperature spread out. This is not difficult but an even simpler problem which gives the same answer is to drop a charged particle onto the horizon and compute the time for the charge to equilibrate. Since the horizon is an electrical conductor the charge density will quickly become uniform.
The basic calculation was done in  \cite{james}. We will redo it here for the ten dimensional case.

The important point that allows an easy calculation is that all thermal
horizons are locally isomorphic to Rindler horizons. Thus consider 10D
flat Rindler space, in Rindler coordinates, $\omega, \r, x^i$.
\bea
ds^2 \eq dx^i  dx^i + dz^2 - dt^2 \cr
\eq dx^i  dx^i + d\rho^2 -\rho^2 d\omega^2
 \label{41}
\eea
where $i = 1,2,...,8$ runs over the directions in the horizon, $z$ is the remaining spatial coordinate, $t$ is Minkowski time\footnote{Note that in this argument $t$ is not the \sc \ time.}, and the Rindler coordinates $\r,\omega$ are defined by
\bea
\rho^2 &=& z^2 - t^2 \cr
z &=& \r \cosh \omega \cr
t &=& \r \sinh \omega.
\label{42}
\eea
The stretched horizon is defined by the surface
\be
\r = l_s
\label{43}
\ee
where $l_s$ is the string length scale, sometimes called $\sqrt{\alpha'}$.

Now let us perturb the stretched horizon by dropping a unit charge  from the point $x^i=0, \ z = 1$. If the charge is initially at rest it will remain at rest in Minkowski coordinates, but in Rindler coordinates it will asymptotically approach the horizon.

Now consider the $\r$ component of the electric field. In fact it is identical to the $z$ component,
\be
E_{\r} = {z -  1 \over \left\{x^i x^i + (z-1)^2 \right\}^{9 \over 2}}.
\label{44}
\ee

According to the membrane paradigm for black holes, the $\r$ component of the electric field on the stretched horizon is identified with the surface charge density, $\sigma$, on the (8D) stretched horizon. Let us rescale the horizon coordinates $x^i$ by defining
\be
x^i = X^i l_s e^{\omega}.
\label{45}
\ee
The surface charge density  $\sigma$, at large Rindler time, is given by
\be
\sigma = {1 \over (l_s e^{\omega})^8}(X^2 +1)^{-{9 \over 2}}.
\label{46}
\ee
It is evident from (\ref{45}) and (\ref{46}) that at Rindler time $\omega$, the charge density has spread to a distance $\Delta x$ given by
\be
\Delta x = l_s e^{\omega}
\label{47}
\ee
Thus we see that the Rindler  time $\omega_*$ needed for the charge to spread over the whole horizon scales like
\be
\omega_* = \log {R_s \over l_s}
\label{48}
\ee
where $R_s$ is the Schwarzschild radius of the horizon.

The Rindler time is connected to the asymptotic observer's time, $t$, by a simple general formula
\be
\omega = {2\pi t \over \beta}
\label{49}
\ee
where $\beta$ is the inverse Hawking temperature. Using (\ref{48}) and
(\ref{49}) we find that the scrambling time is given by
\be
t_* \sim \beta \log {R_s \over l_s}.
\label{410}
\ee

One further suggestive equation is obtained by assuming that $l_s$ is of order the Planck length and recalling that the entropy is a power of
${R_s \over l_p}$.
\be
t_* \sim \beta \log S.
\label{411}
\ee

In terms of the dimensionless variable $\tau = t_* / \beta$
\be
\tau= C \log S
\label{412}
\ee
with $C$ being a constant of order unity (In ordinary units $C$ has dimensions of action and so should be of order $\hbar$). Incidentally, equation (\ref{412}) is very general and applies to ordinary Schwarzschild black holes in all dimensions although the numerical constants may vary.

If we think of the entropy of the black hole as the number of its degrees of freedom then (\ref{412}) shows that black holes are fast scramblers. But we would like to do better and actually identify a concrete hamiltonian fast scrambler. To that end we consider a system of D0-branes in ten-dimensional type 2a string theory.

\subsection{D0-brane black hole}
The string frame metric for the black hole made of the D0-branes 
is~\cite{horowitz} 
\begin{equation}
ds^2 =  -f^{-1/2}\left( 1-{r_H^7 \over r^7}\right)
d t^2  + f^{1/2}
\left[\left( 1-{r_H^7\over r^7}\right)^{-1} dr^2
+ r^2 d\Omega^2_{8}\right]
\label{410a}
\end{equation}
where
\begin{equation}
f = 1 + {r_H^7 \sinh^2 b \over r^7}.
\label{411a}
\end{equation}
Charge (the number of D0-branes) of this black hole is
$n\sim (r_H^7/ g_s \alpha'^{7/2})\sinh 2 b$. 
(We will ignore numerical factors of order one
in this section.)
The parameter $b$ specifies the non-extremality; 
the extremal limit is $b \to \infty$ with $n$ fixed. 

The near-extremal black hole has a dual 
description in terms of supersymmetric quantum
mechanics~\cite{matrix}. More specifically, we consider
the ``decoupling limit'' \cite{IMSY}: 
we let $\alpha'\to 0$ in (\ref{410a}), 
with $U\equiv r/\alpha'$, $U_0\equiv r_H/\alpha'$, 
$g_{YM}^2\equiv g_s \alpha'^{-3/2}$ and $n$ held fixed,
which gives 
\[
 ds^2=\alpha'\left[-\left({g_{\rm YM}^2 n\over U^7}\right)^{-1/2}
\left(1-{U_0^7\over U^7}\right)dt^2
+\left({g_{\rm YM}^2 n\over U^7}\right)^{1/2}
\left\{\left(1-{U_0^7\over U^7}\right)^{-1}dU^2
+ U^2d\Omega_8^2\right\} \right].\nonumber
\]
It is expected that the system is decoupled from
open string excited modes in this limit and can be described by the
super Yang-Mills theory (quantum mechanics) with coupling constant
$g_{\rm YM}$. $U_0$ corresponds to the energy scale that we are 
interested in. 

Energy (mass above extremality) $E$, 
entropy $S$, and the Hawking 
temperature $T$ for the this black hole are~\cite{IMSY}
\begin{equation}
E  \sim  {U_0^7\over g_{\rm YM}^4},\qquad
S \sim  {n^2 U_0^{9/2}\over (g_{\rm YM}^2n)^{3/2}},\qquad
T \sim  {U_0^{5/2}\over (g_{\rm YM}^2 n)^{1/2}},
\label{412a}
\end{equation}
and the dilaton expectation value at the horizon is
\begin{equation}
e^{\phi}\sim {1\over n}\left({g_{\rm YM}^2 n\over U^3}
\right)^{7/4}.  
\end{equation}

Geometry near the horizon is of the Rindler form (\ref{41}). 
(Rindler time $\omega$ is related to the time $t$ by $\omega = 2\pi T t$.)
So we can use the result in the last subsection to find the 
scrambling time. We can trust classical calculation when
the coupling is weak ($e^{\phi}\ll 1$) and the curvature is small 
in string units, which give us the conditions,
$1\ll (g^2_{\rm YM}n/U_0^{3})\ll n^{4/7}$. These are
satisfied in the 't Hooft limit where
we send $n\to \infty$ with 
the effective coupling $g^2_{\rm eff}=g^2_{\rm YM}n/U_0^{3}$ 
fixed.

The linear size of the horizon is 
$R_s\sim (g^2_{\rm YM}n/U_0^{3})^{1/4}\ell_s
\sim (g^2_{\rm YM}n)^{7/10}T^{-3/5}\ell_s$, 
and we find that the scrambling time (\ref{48}) depends on $n$
logarithmically
\footnote{Strictly speaking, in (\ref{419a}),
we cannot scale $n$ arbitrarily with other parameters fixed, 
since it will violate the condition $e^{\phi}\ll 1$. 
We are taking $g_{\rm YM}$ to be very small so that 
$n$ can become large while keeping $e^{\phi}\ll 1$.
We believe this gives the correct large $n$ behavior.}
\begin{equation}
 \omega_*=\log {R_s \over \ell_s}\sim C\log n
\label{419a}
\end{equation} 
with an order one constant $C$. 

For an observer who stays at a radial position $r$ well above 
the stretched horizon, the proper time $t_p$ is related to $\omega$ by
$t_p = (2\pi T)^{-1}\sqrt{g_{00}(r)}\, \omega$. On the other hand
the local proper temperature at $r$ satisfies $\beta(r) = T^{-1}
\sqrt{g_{00}(r)}$. (If we go to infinite distance the temperature 
becomes the Hawking temperature.)
Thus it follows that at any $r$, the scrambling
time $t_*$ is
$$
t_* = {\beta(r)\over 2\pi} \log {R_s \over \ell_s}.
$$
The expression (\ref{419a}) is the scrambling time in unit of
inverse temperature.

As in the case of the Schwarzschild black hole, the
scrambling time is shorter than the time-scale for evaporation 
(i.e. for the approach to extremality). The latter is estimated as 
follows: In unit Rindler time, 
approximately one photon will escape from the potential
barrier outside the black hole 
and is emitted as Hawking radiation~\cite{james}. 
Typical energy of each photon is $T$. Thus the luminosity $L$ 
(energy radiated away 
in unit time in $t$) is of order
\begin{equation}
L=-{dE\over dt}\sim T^2 \sim {g_{\rm YM}^{6/7}E^{5/7}\over n}. 
\end{equation}
From this we find that the evaporation time, 
$t_{\rm evap}\sim n E^{2/7}/g_{\rm YM}^{6/7}$, scales
linearly with $n$, and is larger than the scrambling time 
$t_*\sim T^{-1}\log n\sim n^{1/2}\log n$ in the large $n$ limit. 

We believe the cloning of quantum state will not happen, although
the argument described in Section 1 may seem to have 
a problem at first sight.
The causal structure of the D0-brane black hole is similar
to the one for the Reissner-Nordstrom black hole in four 
spacetime dimensions: Classical geometry of a non-extremal 
black hole has inner and 
outer horizons, and the maximally extended spacetime has 
infinite number of asymptotic regions and timelike singularities.
Geometry near the (outer) horizon is described by 
Rindler space (\ref{41}). The region between outer and inner
horizon is covered by the coordinates of the type
$(X^{+},X^{-})$ in (\ref{01}), with $0\le X^{+}X^{-}\le \infty$.
(Geometry away from the outer horizon is not exactly flat,
and the 2-dimensional part of the metric is of the form 
$f(X^{+}X^{-})dX^{+}dX^{-}$ with
$f(X^{+}X^{-})$ being non-zero except at the inner horizon.) 
Inner horizon is at $X^{+}X^{-}=\infty$, and 
it appears that Bob will have enough time to receive Alice's 
message inside the black hole, which would make the cloning possible
in the thought-experiment in Section 1.

However, the true causal structure for the charged black hole 
is believed to be different from the above one. 
(See~\cite{frolov} for a recent discussion.)
There will be an instability near the inner horizon due to infinite 
blue shift. This instability will produce a spacelike singularity 
at finite $X^{+}X^{-}$. The true causal structure will be 
similar to the one for the Schwarzschild black hole, and we can apply 
the argument for no-cloning in Section 1 also in 
this case.

To interpret the above perturbation in M(atrix) 
theory, note that a perturbation localized on the horizon with
size $\Delta X$ has angular size $\Delta \theta\sim \Delta X/R_s$.
This correspond to the angular momentum of order $\ell\sim 1/\Delta\theta$. 
If we increase the size of the horizon $R_s$ keeping $\Delta X$ fixed, 
the angular momentum scales like $\ell \sim R_s\sim 
(g^2_{\rm YM}n/U_0^{3})^{1/4}$. 
An operator with angular momentum $\ell$ involves at least $\ell$
matrix fields. So we expect that the localized perturbation 
that we studied corresponds to an operator which has
the form of a trace of a product of $\ell$ matrices. 

The unperturbed black hole is a typical pure state with a given energy. 
This state should be a scrambled state in the sense discussed 
in Section 4. Our conjecture is that when we perturb this state by 
the above operator, the system returns to a scrambled state in time
$\log n$ in unit of inverse temperature.

 \setcounter{equation}{0}
\section{ADS/CFT}

Next, let us consider the case of $SU(n)$ Yang Mills theory on a unit 3-sphere. The precise version of the theory that is most easily analyzed is the maximally supersymmetric case. In that case the gravitational dual theory is gravity in $ADS(5) \times S(5)$. The radius of curvature of the ADS space in string units is given by
\be
\R = (g_{\rm YM}^2 n)^{1 \over 4}
\label{51}
\ee

Recall that the ultraviolet regulator of the boundary quantum field theory (QFT) is related to the infrared regulator of the bulk ADS theory \cite{holobound}. The important point for us is that as the regulator length scale in the QFT goes to zero, the proper size of the regulator does not tend to zero in bulk metric units. In fact it is always of order $\R$. In other words a point in the QFT is a large patch of space of size $\R$ in the bulk. Each such patch is described by a system of $n \times n$ matrices, and the entire field theory is a 3D array of such patches.

Let us consider a large ADS black hole of temperature $1/ \beta$ beyond the Hawking Page transition. It is convenient to introduce a field theory UV cutoff with length-scale a few times bigger than (but of order) $\beta$. This will have negligible effect on the system. It also corresponds to replacing the horizon by cells of size $\R$, each of which has $n^2$ degrees of freedom.

Next, we perturb the horizon within one such cell. Typically that will involve perturbing all $n^2$ degrees of freedom. But we wish to excite only a small finite subset. The obvious way to do this is to localize the perturbation on a scale of order $1$ rather than order $\R$.

On scales smaller than $\R$, the horizon of an ADS black hole is of the Rindler form: the curvature of ADS can be ignored. Thus we can use our earlier estimates for the time required to spread over the cutoff patch. Equivalently this is the time for the perturbation to equilibrate among the $n^2$ degrees of freedom. As usual the answer is
of order
\be
\omega_* \sim \log \R.
\label{52}
\ee
In fact we can be a bit more precise. The exact scale invariance of ${\cal{N}}=4$ super Yang Mills requires the coefficient in (\ref{52}) to be proportional to $\beta$.
\be
\omega_* \sim \beta \log \R.
\label{53}
\ee
Using (\ref{51}) we find
\be
\omega_* = C \beta \log n.
\label{54}
\ee
with $C$ being a numerical constant of order $1$.

Once a single cell has been excited a signal will propagate away from that cell with the speed of sound which in the case of a conformal QFT is the speed of light divided by $\sqrt{3}$. 
Evidently, the time needed to scramble the entire sphere is at least
\be
\omega = \sqrt{3}\pi + C \beta \log n.
\label{55}
\ee

Thus we see that in anti de Sitter space scrambling is a combination of fast scrambling on scales smaller than the ADS radius, and slower conventional 3-dimensional scrambling on larger scales.

 \setcounter{equation}{0}
\section{Conclusions}

Naively, black hole horizons are $D-2$ dimensional systems, and as far as counting their entropy they are. But if measured by their scrambling times they are more like infinite dimensional systems. 

It is surprising that a real physical system can scramble that fast. One might have argued that as the number of degrees of freedom increases they have to spread out in space, either along a line, a plane, or in a space-filling way. We can imagine connecting distant degrees of freedom by wires and simulating non-locality, or a higher dimensional system, but eventually the wires will get so dense that there will not be room for more. The fastest scramblers in three spatial dimensions would have a scrambling time of order
$N^{2/3}$. This seems likely to be the case for anything made of ordinary matter. 

But that intuition is wrong when gravity is involved: gravity brings something entirely new into the game, something that looks so non-local that black holes effectively are infinite dimensional.  They are  the fastest scramblers in nature by a wide margin.

The fact that black holes are fast scramblers is not just an interesting curiosity.
The principle of Black Hole complementarity requires that no observer be able to observe
information cloning. This places a bound on how fast Bob can retrieve Alice's q-bit after Alice jumps into the black hole. Before \cite{patrick} the very long retrieval time \cite{Larus}led to a serious ``overkill" situation. Complementarity would have been more compelling if it had just barely escaped inconsistency. A good example is the Heisenberg microscope experiment which not only showed that the Uncertainty Principle \it could not \rm be violated, but that it \it could \rm be saturated.

 The Hayden Preskill experiment together with the fact that black holes are fast scramblers leads to a very gratifying situation:  the retrieval time roughly saturates the complementarity bound derived from un-observability of quantum cloning. Indeed it should be possible to make--and hopefully confirm--a more quantitative version of the bound.

\section*{Acknowledgements}

This paper was inspired by conversations with Patrick Hayden. We are
very grateful to him for explaining the subtleties of quantum
information theory that went into \cite{patrick}. We are also indebted
to Steve Shenker, Joe Polchinski, Debbie Leung, Daniel Gottesman,  and John Preskill for helpful insights and comments.

\end{document}